DeepMetabolism: A Deep Learning System to Predict Phenotype from Genome Sequencing

[a][†]Weihua Guo, [b][†]You (Eric) Xu, [a],[*]Xueyang Feng

[a] Department of Biological Systems Engineering, Virginia Polytechnic Institute and State University (Virginia Tech), VA.
[b] ai.codes, Inc., San Francisco, CA.

[†] W.G. and Y. X. are equally contributed.
[*] To whom correspondence should be addressed: Prof. Xueyang Feng, Department of Biological Systems Engineering, Virginia Polytechnic Institute and State University. Human and Agricultural Biosciences Building I, 1230 Washington St. SW, Virginia Tech, Blacksburg, 24061. Telephone: 540-231-2974, Email: xueyang@vt.edu

**Abstract.** Life science is entering a new era of petabyte-level sequencing data. Converting such "big data" to biological insights represents a huge challenge for computational analysis. To this end, we developed DeepMetabolism, a biology-guided deep learning system to predict cell phenotypes from transcriptomics data. By integrating unsupervised pre-training with supervised training, DeepMetabolism is able to predict phenotypes with high accuracy (PCC>0.92), high speed (<30 min for >100 GB data using a single GPU), and high robustness (tolerate up to 75% noise). We envision DeepMetabolism to bridge the gap between genotype and phenotype and to serve as a springboard for applications in synthetic biology and precision medicine.

**1. Introduction**

High-throughput sequencing technology has brought life science into a "big data" era with an unrivaled explosion of genomic and transcriptomic data[1, 2]. The falling cost (<$1,000 per human genome) and increasing speed (<1 day per human genome) of high-throughput sequencing lead to the snowballing data at petabyte level[3]. However, it is still difficult to transfigure such "Big Data" to valuable biological insights such as cell growth rate and metabolic pathway activities. The gap between genome sequencing and cell phenotypes is one of the biggest challenges for achieving "Data-to-Insight". In recent five years, the rapid development of artificial intelligence, especially deep learning, provides a novel option to overcome this challenge. Deep learning is found to be extremely effective in learning and modeling complex systems based on the graphic processing unit computation[4, 5]. Many deep-learning-based systems such as AlexNet[6] and Deep Speech[7] have developed for applications such as image recognition and speech recognition. Recently, deep-learning-based algorithms such as DeepSEA[8] and DeepChem[9] have also been developed to solve biology-related problems such as sequence alterations[8] and drug discovery[9]. Encouraged by these recent successes, we developed DeepMetabolism, a deep learning system that predicts cell phenotypes from genome sequencing data such as transcriptomics data. In DeepMetabolism, the specific model is designed for *Escherichia coli*, but can be easily extended to other organisms. DeepMetabolism uses biological knowledge to guide the design of neural network structure, and integrates unsupervised pre-training with supervised training for model prediction. As shown in the following sections, DeepMetabolism meets the design criteria of high accuracy, high speed, and high robustness. By doing so, DeepMetabolism can be used to enrich our understanding of genotype-phenotype correlation and be applied in fields of synthetic biology, metabolic engineering and precision medicine.

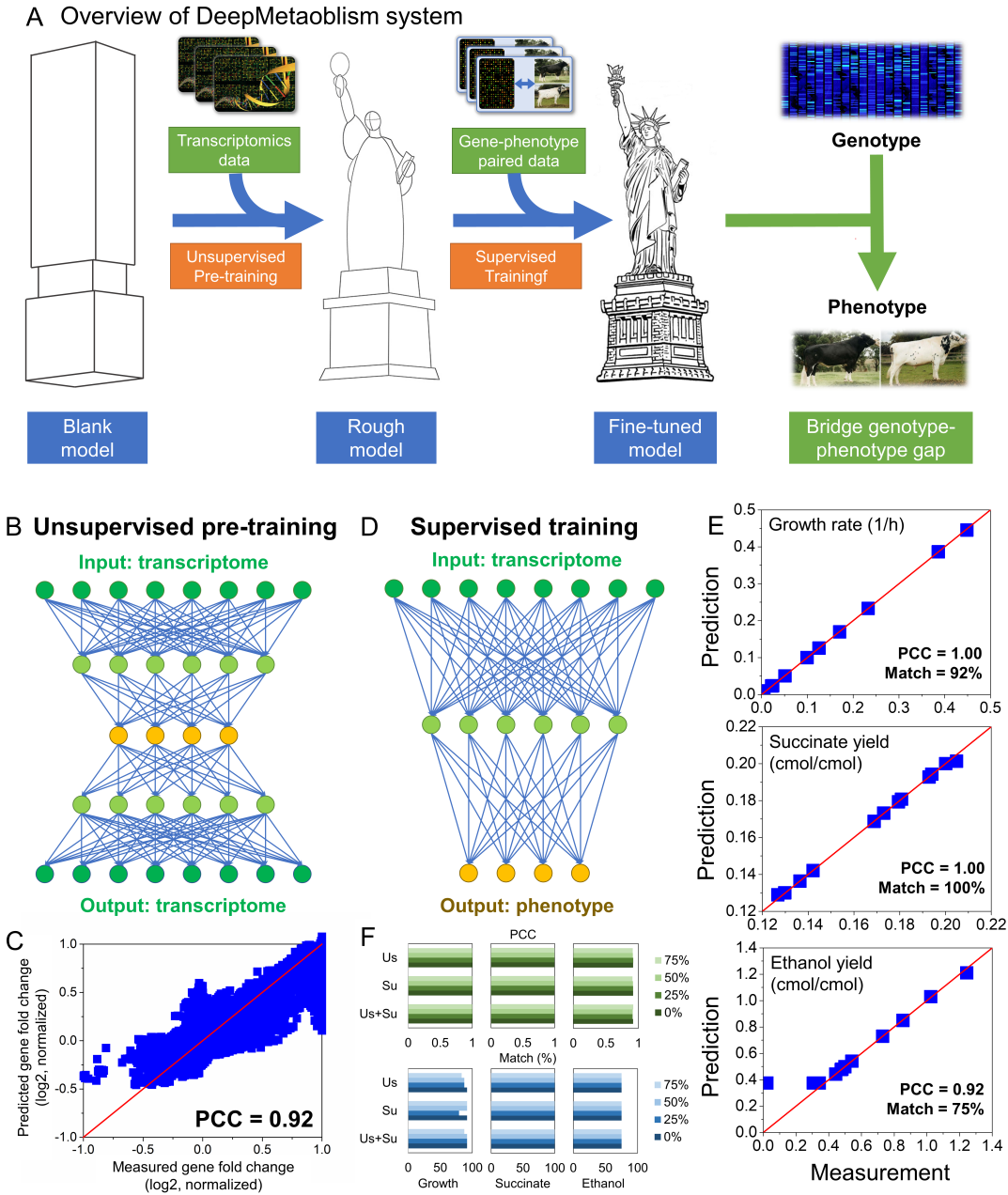

**Figure 1.** Architecture and performance of DeepMetabolism. (A) Overview of DeepMetabolism system. (B) Structure of unsupervised pre-training. Each layer is colored by its corresponding biological definitions. Green represents the gene layer, light green represents the protein layer, yellow represent the phenotype layer. (C) Performance of unsupervised pre-training. Gene expression data was reported in log2(fold-change) and normalized. (D) Structure of supervised training. The meanings of color codes are the same as that in Figure 1B. (E) Performances of supervised training for predicting growth rate (upper panel, unit: 1/h), succinate yield (middle panel, unit: cmol/cmol which is the carbon molar ratio of succinate to glucose), and ethanol yield (lower panel, unit: cmol/cmol which is the carbon molar ratio of ethanol to glucose). (F) Robustness test of DeepMetabolism. Arbitrary noises at different levels (25%, 50%, 75%) were introduced to data used in unsupervised pre-training (Us), supervised training (Su), or both (Us + Su).

## 2. Design of DeepMetabolism

To develop DeepMetabolism, we implemented a two-step, deep neural network model, with two different sets of data. The first step was unsupervised pre-training with transcriptomics data, and the second step was supervised training that used paired data of transcriptomics and phenotype (**Figure 1A**). We expected that unsupervised pre-training would provide a "rough" model that captured the essence of connections between transcriptomics data and phenotype, while supervised training would fine tune this model and lead to highly accurate predictions of phenotype from transcriptomics data.

To design the unsupervised pre-training with transcriptomics data, we built an autoencoder model (**Figure 1B**) with five layers. The first three layers belonged to encoder part, which modeled the connections from gene expressions to phenotype. The last three layers belonged to decoder part, which modeled the connections from phenotype to gene expressions. Each layer in the autoencoder model had unique biological representations. The first layer represented the expression level of 1,366 essential genes of *E. coli* metabolism that were previously reported[10]. The second layer represented the abundance of 1,366 essential proteins of *E. coli* metabolism that were previously reported[10]. The third layer, which was also the "code (bottleneck)" layer of the autoencoder model, represented 110 phenotypes of *E. coli*. The fourth and fifth layers were reconstructed protein layer and reconstructed gene layer, respectively. The nodes in each layer represented corresponding biological components. For example, the 1,366 nodes in the first layer corresponded to 1,366 genes.

In DeepMetabolism, the layers of the autoencoder model were not fully connected. Instead, we applied biological knowledge to rationally define the connections as a strong prior, which could reduce the risk of over-parameterization and increase the training speed. To connect the first layer (i.e., gene layer) and the second layer (i.e., protein layer), we applied the gene-protein association from a well-developed, genome-scale metabolic model of *E. coli*[11] (i.e., iJO1366). To connect the second layer (i.e., protein layer) and the third layer (i.e., phenotype layer), we applied COBRA Toolbox[12] on the genome-scale model iJO1366 to identify the proteins that were essential for a certain phenotype (e.g., proteins that were essential for *E. coli* growth) and connected these proteins with the corresponding phenotype. Totally 1,366 connections were built between the first layer and the second layer and 16,135 connections were built between the second layer and the third layer. The connections between the third layer (i.e., phenotype layer) and the fourth layer (i.e., reconstructed protein layer) mirrored the ones between the second layer (i.e., protein layer) and the third layer (i.e., phenotype layer). Similarly, the connections between the fourth layer (i.e., reconstructed protein layer) and the fifth layer (i.e., reconstructed gene layer) mirrored the ones between the first layer (i.e., gene layer) and the second layer (i.e., protein layer).

This autoencoder model was next trained by using 3,900 transcriptomics profiles that were collected and parsed from Gene Expression Omnibus database[13]. Gene expression data in these transcriptomics profiles was reported as log2(fold change) and normalized to the range in [-1, 1][14]. We shuffled the transcriptomics data for 641 times to generate totally 2,500,000 sets of transcriptomics data. We then trained the autoencoder model using stochastic gradient descending, with a batch size of 5,000 and 500 epochs. We compared the predicted gene expressions by the trained unsupervised model with the input gene expression data to evaluate the model accuracy. As shown in **Figure 1C**, the transcriptomic profiles were reconstructed with high accuracy (Pearson's correlation coefficient, PCC = 0.92).

After unsupervised pre-training, we next designed the supervised training with paired data of transcriptomics and phenotype, and used the same autoencoder model with the first three layers (**Figure 1D**). As connections between layers are regulated by a biological prior, there is a strict one-to-one mapping between the artificial neurons in the phenotype layer and the corresponding phenotype. In supervised training, we trained the network such that the output of the phenotype layer matched the phenotypes we wanted to predict. We trained the model in this study to predict three phenotypes of *E. coli*, namely growth rate ($h^{-1}$), succinate yield (cmol/cmol) and ethanol yield (cmol/cmol), from transcriptomics data. We found that the paired data of transcriptomics and phenotype was extremely rare and the data used in this study was collected from a previously published work[15], which measured the transcriptomics data and cell growth rate, succinate yield and ethanol yield simultaneously at twelve time points during fermentation of an *E. coli* strain. Based on the raw time-dependent data, we used cubic spline smoothing to generate 30,000 synthetic paired data of transcriptomics and phenotype. This allowed us to sufficiently train the model and simultaneously predict multiple phenotypes. We then inherited the autoencoder model from unsupervised pre-training, used it as our initial guess of parameters for supervised learning, and applied 10-fold cross-validation method to evaluate the model prediction. The performance of DeepMetabolism was discussed in the following section.

### 3. Performance of DeepMetabolism

We used two criteria to examine the accuracy of DeepMetabolism: 1) Pearson's correlation coefficient (PCC) between the original observations and corresponding model predictions, and 2) the percentage of matched predictions within 20% error of original observations. As shown in **Figure 1E**, high accuracy was achieved for each of the predicted phenotypes (i.e., growth rate, succinate yield and ethanol yield), with PCC reaching 0.92~1.00 and matched predictions reaching 75~100%. The entire DeepMetabolism system was finished in half an hour (28.3 min for model training and <0.5 min for model prediction) using totally 100 GB data in one GPU (GeForce GTX1080, 7.92 GiB).

We next examined the robustness of DeepMetabolism. One notorious feature of genome sequencing is that the data is often noisy because of numerous reasons such as biological variations[16], platform bias[17], and unreliable human interventions[18]. Therefore, sequence-based algorithms need to be robust to noises in genome sequencing. Here, we introduced artificial noises to the data used in unsupervised pre-training, supervised training, or both. We used a so-called "H-index" method to evaluate the prediction accuracy when using the noisy data. For example, we randomly selected 25% of the data and introduced 25% noises on these selected data, followed by evaluating the PCC and matched predictions. As shown in **Figure 1F**, DeepMetabolism demonstrated high robustness (i.e., almost no decrease in PCC or matched predictions) to various levels of noises (25%, 50% or 75%) no matter which step these noisy data was introduced. Overall, the key features of DeepMetabolism, i.e., high accuracy, high speed, and high robustness, make it a promising tool for mining "Big Data" of genome sequencing.

### 4. DeepMetabolism *vs* other variants

DeepMetabolism is unique because it is a *biology-guided* deep learning system and uses both unsupervised pre-training and supervised training for model prediction. Here, we compared DeepMetabolism (strategy C) with two other alternatives. The first is to change the model structure (strategy A) such that different layers in the model are fully connected. The second (strategy B) is to eliminate unsupervised pre-training while using the same biology-guided network structure. We aimed to answer two questions: *Is biological prior necessary for the success of DeepMetabolism?* and *Is unsupervised pre-training necessary for DeepMetabolism?*

Table 1. Requirements and Performances of Three Deep Learning Systems

| | | Strategy A | Strategy B | Strategy C (DeepMetabolism) |
|---|---|---|---|---|
| Strategy requirement | Biological prior | No | Yes | Yes |
| | Unsupervised pre-training | No | No | Yes |
| | Supervised training | Yes | Yes | Yes |
| Strategy performance | Growth rate PCC | 1.00 | 0.39 | 1.00 |
| | Growth rate Match (%) | 91.7 | 0.10 | 75.0 |
| | Succinate yield PCC | 1.00 | 0.41 | 1.00 |
| | Succinate yield Match (%) | 100.0 | 50.0 | 100.0 |
| | Ethanol yield PCC | 0.92 | 0.67 | 1.00 |
| | Ethanol yield Match (%) | 75.0 | 16.7 | 91.7 |
| | Time (min) | 873.8 | 102.7 | 28.3 |

We first designed supervised training that fully connected each of the three layers as that in DeepMetabolism (i.e., gene layer, protein layer, and phenotype layer). Compared to DeepMetabolism, which only had 17,501 connections, the fully connected supervised learning (strategy A) had over 2 million connections. As shown in **Table 1**, although the prediction accuracy was not jeopardized in the fully connected supervised learning, the time cost dramatically increased from 28.3 min to 873.8 min. Also, although not shown in this study, we do want to point out that the risk of over-parameterization in the fully connected model will increase. In sum, biological knowledge is essential for the performance of DeepMetabolism as it reduces the computational burden and the risk of over-parameterization.

We next designed supervised training that connected in the same way as that in DeepMetabolism (i.e., gene layer, protein layer, and phenotype layer). We tested this supervised training without the guidance of unsupervised pre-training. As shown in Table 1, this "supervised training alone" system (strategy B) failed to achieve the same high accuracy for predictions, with PCC reaching only 0.39~0.67 and matched predictions reaching only 16.7~50.0%. The time cost also increased from 28.3 min to 102.7 min. The increased solution time was due to the large solution space and the lack of "warm start" initial guesses of parameters. Consequently, supervised training alone could not effectively learn the causal relations between transcriptomics and phenotype. In sum, unsupervised pre-training is crucial for DeepMetabolism as it narrows down the solution space of complex genotype-phenotype connections for effective learning.

## 5. Conclusion

DeepMetabolism yields state-of-the-art accuracy, speed and robustness to predict cell phenotypes from genome sequencing. Using biological knowledge to guide model design as well as integrating unsupervised pre-training with supervised training was found to be the key of DeepMetabolism. Future work will focus on extending DeepMetabolism to more complex biological systems for applications such as automated design-build-test cycle of industrial microorganisms for production of fuels and pharmaceuticals as well as sequence-based early diagnosis of metabolic diseases.

**Author contributions.** Y.X. and X.F. contributed to the original idea of DeepMetabolism. W.G. and Y.X. contributed to the data collection and parsing from GEO database. W.G. and X.F. contributed to the design of DeepMetabolism. W.G. and Y.X. contributed to the programming of unsupervised pre-training and supervised training. W.G., Y.X., and X.F. contributed to the computational experiments and result analysis. W.G., Y.X., and X.F. contributed to the manuscript preparation and revision.

**Data and software.** DeepMetabolism is an open source project. All data used in DeepMetabolism, including transcriptomics data for unsupervised pre-training, transcriptomics-phenotype data for supervised training and the genome-scale metabolic model of *E. coli* (iJO1366), as well as the source codes are available at GitHub: https://github.com/gwh120104/deepmetabolism.

**Appendix: Detailed description of DeepMetabolism network.** We designed the DeepMetabolism in the context of *E. coli*. However, the network structure can be extended easily to other organisms. The deep learning system of DeepMetabolism was implemented in Tensorflow[19].

The input layer was normalized log2 gene expression data. The normalization was done as a pre-processing step. The connections between gene expression and protein layer was simplified as a one-to-one linear transformation (each gene maps to one protein and one protein only). The exact linear parameter between gene and protein were trainable parameters that were tuned during the unsupervised pre-training process.

The connections between protein and phenotype layer were regulated by the biological knowledge. For instance, if we denoted the first node in phenotype layer as growth rate, and it was related to 32 proteins in the protein layer, we would only allow this node to be connected with 32 nodes in the protein layer that corresponded to the 32 proteins. Admittedly this was an approximation of the actual biological process, as growth rate would also be related to other factors such as the amount of glucose in the environment. Nevertheless, this biologically regulated network gave us a strong prior in ensuring the network was learning parameters that are biologically motivated.

The relationship between protein and phenotypes was modeled as a nonlinear mapping over the weighted sum of protein level. The nonlinear function we chose was softplus function (tf.nn.softplus). Weights and bias were trainable parameters and were tuned by unsupervised pre-training. Unsupervised pre-training was tuned with AdamOptimizer[20] in TensorFlow by minimizing the sum of squared error between input and reconstructed gene expressions.

Following the unsupervised pre-training, the supervised training had the same input layer, i.e., normalized log2 gene expression data. The same connections between gene expression and protein layer were used in the supervised training as that in the unsupervised pre-training. The linear parameters between gene and protein layers that were obtained from the unsupervised pre-training process were used in the supervised training process.

The same connections between protein and phenotype layer were used in the supervised training as that in the unsupervised pre-training. The weights of nonlinear mapping between protein and phenotype layers that were obtained from the unsupervised pre-training process were used as initial values in the supervised training process. To fine tune these nonlinear weights, we chose the nonlinear function as softplus and tuned supervised training process with AdamOptimizer[20] in TensorFlow by minimizing the summed relative errors (absolute values) between observed and predicted phenotypes.

**References**
1. Costa, F.F. Big data in genomics: challenges and solutions. *GIT Lab J* **11**, 1-4 (2012).
2. Ward, R.M., Schmieder, R., Highnam, G. & Mittelman, D. Big data challenges and opportunities in high-throughput sequencing. *Systems Biomedicine* **1**, 29-34 (2013).
3. Eisenstein, M. Big data: The power of petabytes. *Nature* **527**, S2-S4 (2015).
4. LeCun, Y., Bengio, Y. & Hinton, G. Deep learning. *Nature* **521**, 436-444 (2015).
5. Schmidhuber, J. Deep learning in neural networks: An overview. *Neural Networks* **61**, 85-117 (2015).
6. Krizhevsky, A., Sutskever, I. & Hinton, G.E. in Advances in neural information processing systems 1097-1105 (2012).